\documentclass[]{spie}  

\usepackage{amsmath,amsfonts,amssymb}
\usepackage{graphicx}
\usepackage[colorlinks=true, linkcolor=blue, filecolor=blue, citecolor=blue, urlcolor=blue]{hyperref}    
\usepackage{siunitx}

\title{The Survey of Surveys: machine learning for stellar parametrization}

\author[a]{Turchi, A.}
\author[a,b]{Pancino, E.}
\author[a]{Rossi, F.}
\author[c]{Avdeeva, A.}
\author[d,b]{Marrese, P.}
\author[d,b]{Marinoni, S.}
\author[a]{Sanna, N.}
\author[a]{Tsantaki, M.}
\author[b]{Fanari, G.}

\affil[a]{INAF-Osservatorio Astrofisico di Arcetri, L.go Enrico Fermi 5, Firenze, Italy}
\affil[b]{Space Science Data Center, Via del Politecnico SNC, I-00133 Rome, Italy}
\affil[c]{Institute of Astronomy, Russian Academy of Sciences, 48 Pyatnitskaya St., Moscow 119017, Russia}
\affil[d]{INAF -- Osservatorio Astronomico di Roma, Via Frascati 33, 00040, Monte Porzio Catone, Roma, Italy}

\authorinfo{Send correspondence to Alessio Turchi - e-mail: alessio.turchi@inaf.it}

\begin{document} 
\maketitle

\begin{abstract}
We present a machine learning method to assign stellar parameters (temperature, surface gravity, metallicity) to the photometric data of large photometric surveys such as SDSS and SKYMAPPER. The method makes use of our previous effort in homogenizing and recalibrating spectroscopic data from surveys like APOGEE, GALAH, or LAMOST into a single catalog, which is used to inform a neural network. We obtain spectroscopic-quality parameters for millions of stars that have only been observed photometrically. The typical uncertainties are of the order of 100K in temperature, 0.1 dex in surface gravity, and 0.1 dex in metallicity and the method performs well down to low metallicity, were obtaining reliable results is known to be difficult.
\end{abstract}


\section{INTRODUCTION}
\label{sec:intro} 
In the last few years large spectroscopic surveys provided a huge amount of photometric measurements for hundreds of millions (up to billions) of stars,  either at low-medium resolution such as the RAdial Velocity Experiment (RAVE) [\citenum{rave}], the Sloan Extension for Galactic Understanding and Exploration (SEGUE) [\citenum{segue}] and the Large sky Area Multi Object fiber Spectroscopic Telescope (LAMOST) [\citenum{lamost}], or at high-resolution such as the Galactic Archaeology with HERMES (GALAH) [\citenum{galah}], the Apache Point Observatory Galactic Evolution Experiment (APOGEE) [\citenum{apogee}] and the Gaia-ESO survey [\citenum{gaiaesosyrvey}]. The data provided by these surveys allow to give a precise estimates of key parameters such as effective temperature (T${\rm{eff}}$), surface gravity (log\,$g$) and iron metallicity ([Fe/H]) for few millions of stars in the Milky Way. 
The availability of high-quality spectroscopic measurement, together with a good estimation of distance and reddening, is of paramount importance to derive high quality estimates of the above parameters from photometric surveys.\\
In recent years, these surveys spawned a many works focused on Machine Learning (ML) methods (i.e. Neural Networks or simpler methods), such as the Cannon [\citenum{cannon}], the Payne [\citenum{payne}] and StarNet [\citenum{starnet}]. ML saw a huge development in the last decades of XX century and rose to a widespread usage in the first decades of XXI century. The term can be used as a general hat to cover different disciplines from Artificial Intelligence to Neural Networks and Computational Statistics. In general we refer to ML techniques when based on algorithms that make use of heterogeneous data to automatically ``learn'' and build a  ``model'' that is used to produce a desired output, using statistical methods.\\
ML methods applied to spectroscopic star catalogues mainly focus on the analysis of the provided data to many different purposes, i.e. trying to enhance the physical model used to compute the parameters, thus this field is already mature in the case where the previous parameters are directly derived from spectroscopic data.\\
However spectroscopic quality measurements are hard to perform on large stellar samples, requiring a lot of telescopes time and the scientific community is thus lacking high-quality estimates on large stellar samples, thus limiting the impact of studies that use the ML-derived spectroscopic data.\\
There are however many photometric surveys that include hundreds of millions (up to billions) of stars that provide observations in many different bands, such as the Sloan Digital Sky Survey (SDSS) [\citenum{sdsssurvey}], the SkyMapper Southern (SMSS) [\citenum{smsssurvey}] and the Two Micron All-Sky Survey (2MASS) [\citenum{2masssurvey}]. While not as accurate as spectroscopic data, photometry can indeed be used to derive a rough estimate of T${\rm{eff}}$, log\,$g$ and [Fe/H], and is used by many astronomers to analyze huge star samples.\\
With this in mind, in this contribution we provide a first preliminary ML application that aims to enhance low-quality photometric measurements and thus improve the accuracy from pre-determined analytic estimates of the star parameters. The strength of this method is that we do not try to produce estimates from scratch, thus we are less impacted by large deviations of the model from the real measurement. To achieve this goal and train our ML model, we used the spectroscopic estimates of  T${\rm{eff}}$, log\,$g$ and [Fe/H] provided by the SoS catalogue [\citenum{tsantaki22}], which is a critical compilation of spectroscopic parameters from survey data. Also this allow us to make use of the astrometric, photometric, and spectroscopic data from the Gaia mission, combined with large photometric datasets such as SkyMapper Southern Survey (SMSS), SDSS and others that provide us huge samples with hundreds of millions of stars, allowing us to maximize the applicability of the method itself.\\

\section{Input catalogs and training set}
\label{sec:ml}

Our neural network (NN) model was designed to produce survey-quality values for effective temperature (Teff), surface gravity (log g), and iron metallicity ([Fe/H]) with spectroscopic-like accuracy using solely standard photometric data.

Initially, we integrated Gaia data (Gaia Collaboration et al. 2016) with Gaia distances (Bailer-Jones et al. 2021). This was supplemented by data from the SkyMapper Southern Survey (SMSS, Keller et al. 2007), forming a comprehensive input dataset that encompass much of the visible spectrum.
We performed a set of training cycles utilizing the NN framework. The supervised training methodology was guided by the 'true' values for Teff, log g, and [Fe/H], which the NN sought to accurately predict.\\

\subsection{Input dataset}

We started to build our input dataset from the {\em Gaia} DR3 dataset\footnote{\url{https://gea.esac.esa.int/archive/}} [\citenum{gdr3}]. We used the software from Marrese et. al. [\citenum{marrese2017,marrese2019}] to perform a precise cross-matching between catalogs, a process greatly enhanced by the detailed proper motions available, with all coordinates conforming to the {\em Gaia} DR3 system. For sample purification, we applied metrics such as stellar blending, binarity, non-stellarity, and variability from the {\em Gaia} catalog. During the neural network's training phase, we leveraged data from the catalog including magnitudes, colors, and other stellar parameters, aiming to enrich the data input and improve the precision and accuracy of our results. Crucially, we incorporated distances derived by [\citenum{bailer21}], which are critical for accurately determining log\,$g$ and [Fe/H] values.\\

We applied several selection criteria to {\em Gaia DR3} catalog in order to start from very clean data in this preliminary NN test:
\begin{itemize}
    \item{sources lacking either {\tt phot\_bp\_mean\_mag} or {\tt phot\_rp\_} {\tt mean\_mag} were removed;}
    \item{sources with {\tt ipd\_frac\_multi\_peak\,$>$\,10} or {\tt ipd\_frac\_} {\tt odd\_win\,$>$\,10} were removed, to avoid disturbance by neighboring objects [\citenum{mannucci22}], as well as sources with {\tt ruwe\,$>$\,1.4} [\citenum{lindegren21}];}
    \item{sources with the {\tt non\_single\_star} or {\tt VARIABLE} flags were removed, as well as those with the {\tt in\_qso\_candidates} and {\tt in\_galaxy\_candidates} flags;}
    \item{we only included sources having a distance in [\citenum{bailer21}] and in particular we decided to use the geometric distance determinations because we noted, a posteriori, that they produced slightly better results than the photo-geometric ones};
    \item{we removed stars with a spectroscopic rotational broadening ({\tt vbroad}) of more than 30\,km\,s$^{-1}$, for the few stars for which this parameter was available, because these stars can have altered colors and less reliable spectroscopic parameters;}
    \item{we removed all stars with a photometric temperature (teff\_gspphot) greater than 7500K;}
    \item{finally, we also removed stars with G\,$>$\,18\,mag, parallax\_error\,$>$\,0.1 or astrometric\_sigma5d\_max\,$>$\,0.1, which have worse astrometric quality parameters, because they do not always allow for a reliable distance determination. After the first tests, in fact, we noticed that $\simeq$\,13\% of the stars in the training sample had conflicting properties in different catalogs. In the spectroscopic surveys they were clearly giants, while they appeared to lie on the main sequence of the absolute and dereddened {\em Gaia} color-magnitude diagram. This conflicting information confused the algorithm, providing wrong log\,$g$ determinations for a large part of the training set. Therefore, for this experiment, we decided to use the cleanest possible sample. The adopted cut reduced the stars with conflicting information to about 1\%.}
\end{itemize}

As a result, we pre-selected an initial sample of almost 27 million stars from the {\em Gaia} DR3 catalogue, that we further selected as described in the following.\\

We then moved to integrate the SkyMapper DR2 dataset\footnote{\url{https://skymapper.anu.edu.au/table-browser/dr2/}} [\citenum{huang21}], which offers photometry in six {\em uvgriz} bands for approximately half a billion stars. We performed a cross-matching of SMSS data with a selected set of sources from {\em Gaia} DR3, employing algorithms developed by [\citenum{marrese17,marrese19}]. We eliminated all stars in each catalog that corresponded to multiple entries in the other, keeping only the cleanest sources.\\

To filter out non-stellar objects from the SkyMapper catalog, we applied the criterion $classStar>0.8$ and excluded problematic data using the conditions $flag=0$ and $flagPSF=0$. Stars with any magnitudes exceeding 25\,mag were removed. Following these filtering steps and cross-matching with the refined {\em Gaia} DR3 sample, we retained a significant sample of nearly 11.4 million stars, spatially distributed in the southern hemisphere. In subsequent work we plan to add other surveys to cover the Northern hemisphere as well. In a future paper we already planned to add other surveys (e.g. SDSS) to cover also the northern hemisphere and expand the dataset.

The absolute magnitudes were computed using the distances from {\em Gaia} ($D$), according to the formula:
\begin{equation}
\label{eq:absmag}
M_{abs}(B)= M_{rel}(B) - 5*\log_{10}(D) + 5 - A(B)
\end{equation}
where $M(B)$ indicates the magnitudes, whether absolute or relative, for each band $B$. The extinction coefficients ($A(B)$) were derived from $E(B-V)$ extinction maps, indicated by the parameter ebmv\_sfd, and adjusted using the transformation coefficient for each of the SMSS bands\footnote{\url{https://skymapper.anu.edu.au/filter-transformations/}}.\\

In table \ref{tab:gsk_in} we list all the parameters selected to be used as an input to the NN model.\\
\begin{table}
\caption{Input parameters used for the NN model.}
\label{tab:gsk_in}
\centering                         
\begin{tabular}{ll}        
\hline\hline                
Parameter   &   Description\\
\hline  
r\_med\_geo   &   geometric distance from [\citenum{bailer21}]\\
phot\_bp\_mean\_mag    &   Integrated BP mean magnitude from Gaia\\
phot\_rp\_mean\_mag &   Integrated RP mean magnitude from Gaia\\
Uabs    &  Absolute magnitude in U band\\
Uabs\_nored   &  Absolute magnitude in U band excluding reddening \\
Vabs    &  Absolute magnitude in V band\\
Vabs\_nored   &  Absolute magnitude in V band excluding reddening \\
Gabs    &  Absolute magnitude in G band\\
Gabs\_nored   &  Absolute magnitude in G band excluding reddening \\
Rabs    &  Absolute magnitude in R band\\
Rabs\_nored   &  Absolute magnitude in R band excluding reddening \\
Iabs    &  Absolute magnitude in I band\\
Iabs\_nored   &  Absolute magnitude in I band excluding reddening\\
Zabs    &  Absolute magnitude in Z band\\
Zabs\_nored   &  Absolute magnitude in Z band excluding reddening \\
e\_u\_psf &   Error in U-band PSF magnitude\\
e\_v\_psf &   Error in V-band PSF magnitude\\
e\_g\_psf &   Error in G-band PSF magnitude\\
e\_r\_psf &   Error in R-band PSF magnitude\\
e\_i\_psf &   Error in I-band PSF magnitude\\
e\_z\_psf &   Error in Z-band PSF magnitude\\
teff\_gspphot   &   T$_{\rm{eff}}$ from GSP-Phot Aeneas using BP/RP spectra \\
mh\_gspphot  &   [Fe/H] from GSP-Phot Aeneas best library using BP/RP spectra\\
logg\_gspphot    &   log\,$g$ from GSP-Phot Aeneas best library using BP/RP spectra\\
teff\_gspphot\_err    &   Error on T$_{\rm{eff}}$\\
logg\_gspphot\_err  &   Error on log\,$g$\\
mh\_gspphot\_err    &   Error on [Fe/H]\\
\hline 
\end{tabular}
\end{table}

\subsection{Training data}
\label{sec:specsos}

Our star sample with known spectroscopic parameters—T${\rm{eff}}$, log\,$g$, and [Fe/H]—was derived from the first SoS data release [\citenum{tsantaki22}]\footnote{\url{http://gaiaportal.ssdc.asi.it/SoS/query/form}}. This release, primarily featuring radial velocities (RV), also included an early version of a stellar parameter catalog that we will call SoS-spectro, for brevity. This catalog was utilized to examine the influence of various parameters on RV trends across different surveys and incorporated data from APOGEE DR16 [\citenum{ahumada20}], GALAH DR2 [\citenum{buder18}], Gaia-ESO DR3 [\citenum{gilmore12}], RAVE DR6 [\citenum{steinmetz20a, steinmetz20b}], and LAMOST DR5 [\citenum{deng12}]. Despite the straightforward nature of its homogenization method [\citenum{tsantaki22}], SoS-Spectro displays typical uncertainties around $\lesssim$100,K for T${\rm{eff}}$ and about $\simeq$0.1,dex for log\,$g$ and [Fe/H], encompassing nearly 5.5 million stars. SoS-Spectro has been effectively used to characterize both Landolt and Stetson secondary standard stars [\citenum{pancino22}], providing reliable results for challenging parameters such as log\,$g$ and [Fe/H].\\

Finally, after cross-matching the SoS Catalogue with our previously collected input samples, we developed a new datasets with 624410 stars. These contain the spectroscopic measurements for T$_{\rm{eff}}$, log\,$g$, and [Fe/H], which will be pivotal for the reference outputs during the machine learning algorithm’s training and testing phases.

\section{Neural network model}
\label{sec:nn}

Prior to the research presented in this paper, we evaluated several machine learning (ML) and deep learning (DL) architectures to identify the most effective approach. In previous work [\citenum{pancino22}], we tested basic machine learning algorithms such as Random Forest (RF), K-Neighbours, and Support Vector Regression (SVR) on a relatively small dataset of approximately 6000 stars from Landolt and Stetson photometry. From this evaluation, SVR emerged as more effective than more complex deep learning techniques like Multilayer Perceptron (MLP) networks, despite its simplicity. SVR, a linear model used for both regression and classification, operates by identifying a hyperplane that maximizes the margin between the support vectors while minimizing the regression error, effectively separating the N-dimensional data points onto an N-1 dimensional hyperplane, which then acts as the regression function.

Conversely, MLP utilizes a nonlinear approach involving multiple layers of artificial neurons, each with specific activation functions, making it part of the broader DL category. This network includes an input layer that receives data vectors, several hidden layers, and an output layer that generates the regression results. Each neuron across the layers is linked to all inputs, with weights that influence the activation of the neuron based on a nonlinear activation function. MLP networks are trained using a dataset to fine-tune these weights to minimize output errors compared to known true values, as determined by a loss function (e.g., mean error, RMSE).

Despite MLP's ability to handle larger datasets more effectively as they often perform better with increased data volume, they are less interpretable and considered ``black boxes'' because it is challenging to discern how each layer's calculations contribute to the final outcomes. Additionally, MLPs are prone to overfitting, which means they might perform excellently on training data but generalize poorly to new datasets. To combat this, techniques such as weight regularization and dropout layers are employed to prevent over-strong connections and enhance network robustness by randomly deactivating neurons during training (see section \ref{secimp}).

\subsection{Implementation}
\label{secimp}
For this study, we adopted an MLP network architecture using the Keras Python interface for the TensorFlow library. After establishing a clean, cross-matched training and testing catalogue (see section \ref{sectest}), the entire training process was executed on a standard desktop PC, with each session taking only a few hours of CPU time.

The MLP was constructed with sequential layers: a fully connected (dense) layer with a Leaky Rectified Linear Unit (Leaky ReLU) activation function for improved performance and convergence, followed by a batch normalization layer to stabilize the mean output near zero with a standard deviation of one, and a dropout layer to prevent overfitting. This setup is particularly crucial given the potential noise in the dataset, which could lead to overfitting rather than learning from the actual physical properties.

Our network design included 18 hidden layers of varying sizes, from 80 to 160 elements, arranged in a diamond shape to optimize processing. During the initial phase of training, we minimized the mean absolute error to address the maximum error across the dataset and mitigate the impact of outliers. For the subsequent phases, we adjusted to minimize the symmetric mean absolute percentage error (SMAPE), defined by the equation:
\begin{equation}
    f(y,y_{pred})=\frac{2*\lvert y-y_{pred} \rvert}{\lvert y \rvert + \lvert y_{pred} \rvert + \epsilon}
\end{equation}

where $y$ is the actual value, $y_{pred}$ is the predicted value, and $\epsilon=10^{-3}$ prevents function explosion.

Each network was tailored to predict specific parameters—T$_{\rm{eff}}$, log\,$g$, and [Fe/H]—with a single-element wide output layer for each. Additional strategies were implemented to handle missing data effectively, ensuring the network could still operate robustly when encountering NaN values in large catalogues.

\section{Training and Testing}
\label{sectest}
During the training of the MLP model we allocated 80\% of the data for training, 10\% for internal validation, and the remaining 10\% for final testing. These subsets were randomly selected, and the networks were run multiple times to ensure consistent and repeatable results across different test subsamples.

\begin{figure}
\centering
\includegraphics[width=1.0\textwidth]{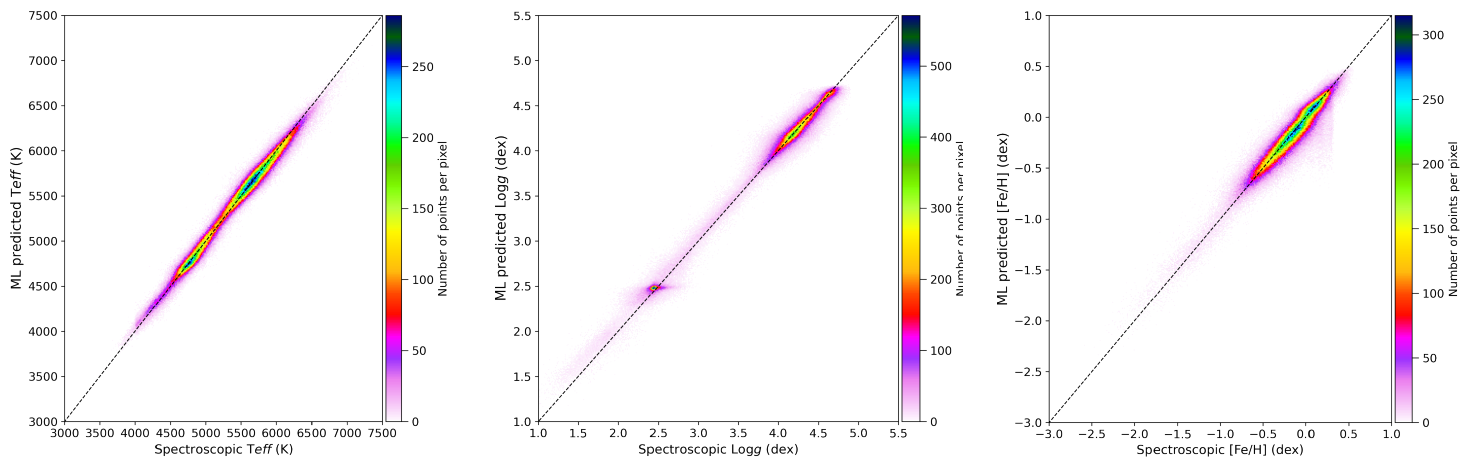}
\caption{Results of the MLP model performance on the test dataset. From left to right we show the scatter plots for T$_{\rm{eff}}$, log\,$g$, and [Fe/H]. The color indicates the density of data points, while the 1:1 relation is plotted as a dashed line.}
\label{fig:resultstest}
\end{figure}

During the training, the network cycles through the dataset multiple times in what are called epochs, each time adjusting its internal parameters (or coefficients) to minimize the loss function tested against the validation subset, which is not used for training. The goal is to align the network's output as closely as possible with the known spectroscopic values (T$_{\rm{eff}}$, log\,$g$, and [Fe/H]) during training. Once the network achieves satisfactory performance or runs for a predetermined number of epochs, its final performance is evaluated on the test subset, which is entirely separate from the training and validation data, to verify the accuracy of the network's output against the desired parameters post-training.

It is important to recognize that any significant biases or deviations in the test subset's T$_{\rm{eff}}$, log\,$g$, and [Fe/H] measurements will similarly influence the neural network's output, as the machine learning algorithm's accuracy cannot exceed that of the reference measurements used during training. If this happens this is a clear sign of overfitting.\\

In Fig. \ref{fig:resultstest}, we compare the spectroscopic measurements from SoS with the ML predictions on the test subsample for each parameter. To quantify the model's performance, we present both the mean and median errors in Tab. \ref{tab:statsres} to identify potential biases in the ML predictions. We also provide the standard deviation and the Median Absolute Deviation (MAD), defined as:
\begin{equation}
\label{eq:mad}
    MAD(x)=median(|x-median(x)|)*k
\end{equation}
where $k=1.4826$ is the normalization constant to compare MAD to standard deviation.\\

\begin{figure}[ht]
\centering
\includegraphics[width=1.0\textwidth]{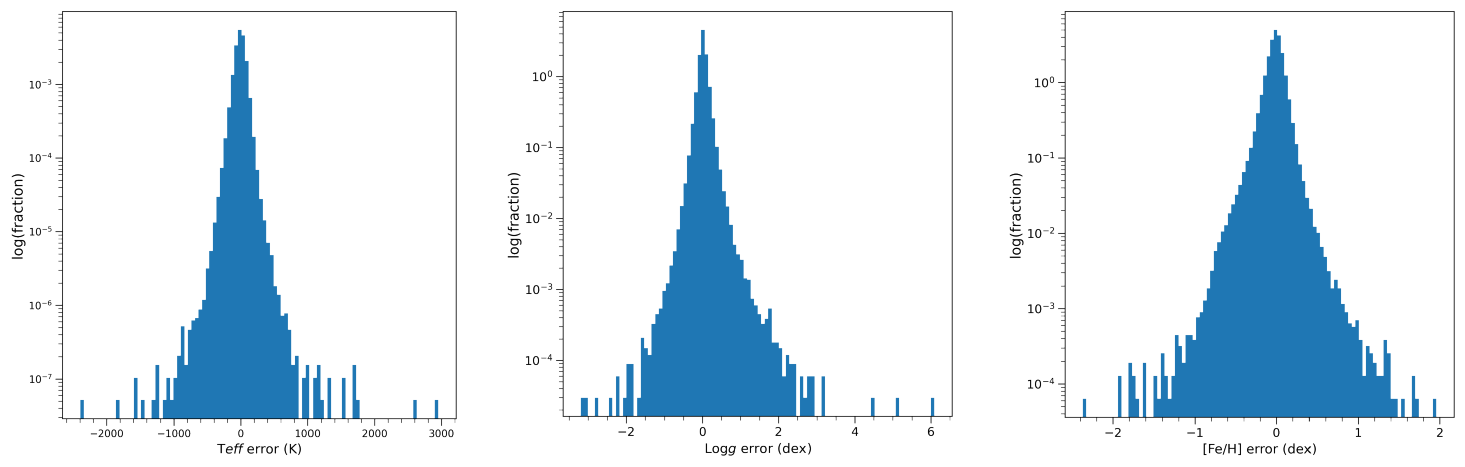}
\caption{Distribution of ML prediction errors on the test dataset, computed as the difference between ML-predicted value and SoS value. The y scale is the logarithm of the number of samples in each bin, rescaled by the total sample size.}
\label{fig:disterror}
\end{figure}

\begin{table}[ht]
    \centering
    \caption{Errors of the ML-predicted variables on the test sample for T$_{\rm{eff}}$, log\,$g$, and [Fe/H], with respect to the SoS spectroscopic ones.}
    \label{tab:statsres}
    \begin{tabular}{lccc}
    \hline
    \hline
        & T$_{\rm{eff}}$ (K) & \textbf{log\,$g$} (dex) & [Fe/H] (dex) \\
        \hline
        Mean & --6.8 & \num{0.016} & \num{0.016} \\
        Median & -5.0 & \num{0.011} & \num{0.0097} \\
        St.Dev & 85.0 & \num{0.14} & 0.12 \\
        MAD & 70.5 & \num{0.086} & \num{0.083} \\
        \hline
        \hline
    \end{tabular}
\end{table}

Our findings indicate that the results of this initial implementation of the MLP model are quite accurate. Predicted T$_{\rm{eff}}$ aligns with the SoS spectroscopic measurements within a 80-85 K range, depending on the statistical measure used. Log\,$g$ accuracy is within 0.14-0.09 dex, and the error for [Fe/H] predictions ranges from 0.12-0.08 dex. The residual bias on the parameters, when compared with standard deviation, is completely irrelevant. We are still trying to understand if we can further improve the NN model to improve performance on the rare outliers.\\

Although the average and median performances are very reliable, it's noteworthy that the disparity between the standard deviation and the MAD suggests occasional large errors by the model. In Fig.\ref{fig:disterror}, we illustrate the complete error distribution on a logarithmic y-scale, highlighting these extremely rare but largely discrepant results.

\section{Final sample}

\section{Conclusions}
\label{sec:end}
In this contribution we presented a preliminary study that shows the feasibility of using an ML model trained on high-quality spectroscopic data to derive key stars parameters, such as T${\rm{eff}}$, log\,$g$ and [Fe/H], from lower quality photometric data. The performance of this method applied to photometric data from Gaia and SMSS merged catalogues shows extremely good performance in determining estimates for the above parameters, with errors (on the test set) characterized by a standard deviation of 85.0~$K$ for T${\rm{eff}}$, 0.14 $dex$ for log\,$g$ and 0.12 $dex$ for [Fe/H]. We managed to obtain this thanks to the high-quality spectroscopic measurements provided by the SoS catalogue which were used for the training. This allow us to apply the method to a potentially huge star datasets with hundreds of millions and up to billions of stars, thus providing to the astronomic community the largest high-quality data sample up to date. Future improvements of this work will include an extension to datasets covering also the northern hemishpere (such as SDSS) and a more detailed statistical analysis on the ML predictions outside the test set, where no spectroscopic-quality reference measurements are available. A comparison with other sources of high-quality spectroscopic measurements will also allow us either to improve or to better characterize the model performance.

\acknowledgments
Co-funded by the European Union (ERC, StarDance, GA 101093572, PI E.~Pancino). Views and opinions expressed are however those of the author(s) only and do not necessarily reflect those of the European Union or the European Research Council. Neither the European Union nor the granting authority can be held responsible for them.

\end{document}